\documentclass[]{aa}
\usepackage{txfonts}
\usepackage{times}
\usepackage{natbib}
\usepackage{graphicx}
\usepackage{epsfig,longtable,lscape}
\usepackage{epsfig}
\usepackage{rotating}
\usepackage{hyperref}
\bibpunct[]{(}{)}{;}{a}{}{,}

\begin{document}
	
\def\pdot {\dot P}
\def\nudot {\dot \nu}
\def\Omdot {\dot \Omega}
\def\ltsima{$\; \buildrel < \over \sim \;$}
\def\lsim{\lower.5ex\hbox{\ltsima}}
\def\gtsima{$\; \buildrel > \over \sim \;$}
\def\gsim{\lower.5ex\hbox{\gtsima}}
\def\msun{~M_{\odot}}
\def\rsun{~R_{\odot}}
\def\mdot {\dot M}
\def\XMM{{\em XMM--Newton}}
\def\SAX{{\em BeppoSAX}}
\def\Swift{{\em Swift}}
\def\RXTE{{\em RXTE}}
\def\Astro{{\em AstroSat}}
\def\four{{4U 0728-25}}
\def\EPIC{{\em EPIC}}
\def\MOS{{\em MOS}}
\def\pn{{\em pn}}
\def\RGS{{\em RGS}}
\def\lx{$L_{\rm X}$}
\def\fx{$f_{\rm X}$}
\def\flux{\mbox{erg cm$^{-2}$ s$^{-1}$}}
\def\lum{\mbox{erg s$^{-1}$}}
\def\countsec{\hbox{counts s$^{-1}$}}
\def\fph{ph cm$^{-2}$ s$^{-1}$}
\def\nh{$N_{\rm H}$}
\def\chisqnu {$\chi^{2}_{\nu}$}

\title{The persistent nature of the Be X-ray binary pulsar \four}

\author{N.~La~Palombara\inst{1}, L.~Sidoli\inst{1}, S.~Mereghetti\inst{1}, P.~Esposito\inst{1,2}, G. L. Israel\inst{3}}

\institute{INAF--Istituto di Astrofisica Spaziale e Fisica Cosmica di Milano, 
	via A. Corti 12, 20133 Milano, Italy\\
	e-mail: \href{mailto:nicola.lapalombara@inaf.it}{nicola.lapalombara@inaf.it}
	\and Scuola Universitaria Superiore IUSS Pavia, Palazzo del Broletto, piazza della Vittoria 15, 27100 Pavia, Italy
	\and INAF--Osservatorio Astronomico di Roma, via Frascati 33, 00040 Monteporzio Catone, Italy}

\date{Received DD Month YYYY; accepted DD Month YYYY}

\titlerunning{Persistent nature of \four}

\authorrunning{N.~La~Palombara et al.}

\abstract{We report the results obtained with a \XMM\ observation, performed in April 2023, of the poorly known Galactic Be X-ray binary pulsar \four. It was revealed at a flux level (not corrected for the absorption) \fx(0.2-12 keV) = 1.7$\times 10^{-11}$ \flux, which implies an unabsorbed source luminosity \lx $\simeq 1.3 \times 10^{35}$ \lum: this is the minimum luminosity ever observed for this source. We measured a pulse period $P_{\rm spin}$ = 103.301(5) s, a value $\simeq$ 0.15 \% longer than that estimated in 2016 with \Astro. The pulse profile shows a broad single peak at all energies, with a limited energy dependence and a small increase in the pulsed fraction with energy. The time-averaged \textit{EPIC} spectrum can be described equally well by four different emission models, either with a single non-thermal component (a partially covered power law or a cut-off power law), or with a thermal component in addition to the non-thermal one (a black body plus a power law, or a collisionally ionised gas plus a cut-off power law). All of them provided an equally good fit and, in the case of the power--law plus black--body model, the thermal component is characterized by a high temperature ($kT_{\rm BB} \simeq$ 1.5 keV) and a small size ($R_{\rm BB} \simeq$ 240 m), comparable with that of the neutron-star polar caps. A spectral variability along the pulse phase is present, which suggests a flux variation of the black-body component. These results show that, for its luminosity level, flux variabilty over long time scales, and spectral properties, \four\ is very similar to most of the persistent Be X-ray binaries. Therefore, it can be considered a member of this class of sources.

}

\maketitle

	\section{Introduction}   

Although most Be X-ray binaries (BeXRBs) are transient sources composed of a neutron star (NS) in a wide and eccentric orbit (with $e > 0.3$) around the Be star, a growing class of BeXRBs is that of the persistent Be/NS binaries \citep{ReigRoche99}, which are characterized by a rather constant and low luminosity ($L_{\rm X} \sim 10^{34-35}$ \lum) and a long spin period ($P_{\rm spin} >$ 100 s). These properties suggest that the NS orbit around the Be star is rather wide ($P_{\rm orb} \gsim$ 30\,days) and nearly circular ($e < 0.2$, \citet{Pfahl2002}). Therefore, it is continuously accreting material from the low--density regions of the fast polar wind of the companion star.

The first sample of these sources was composed of the four sources \mbox{X Persei} (which is considered the prototypical persistent BeXRB), \mbox{RX J0146.9+6121}, \mbox{RX J1037.5-5647}, and \mbox{RX J0440.9+4431}. In recent years, the observations performed with the most sensitive X-ray observatories led to the identification of other confirmed or candidate sources of this type, such as \mbox{Swift J045106.8-694803} \citep{Bartlett+13}, \mbox{CXOU J225355.1+624336} \citep{LaPalombara+21}, \mbox{XTE J1906+090} \citep{Sguera+23}, and \mbox{4XMM J182531.5-144036} \citep{Mason+24}. This suggests that this class of X-Persei–like persistent BeXRBs could constitute a significant part of the population of Galactic unidentified low-luminosity X-ray sources.

An additional candidate member of the persistent BeXRBs is the Galactic Be pulsar \four, which is a poorly studied X-ray source. It was discovered with \textit{Uhuru} \citep{Forman+78} and observed also with \textit{Ariel V} \citep{Warwick+81} and \textit{HEAO-1} \citep{Wood+84}. The optical counterpart, V441 Pup, is at an estimated distance (based on \textit{Gaia EDR3} data) $d = 7.6^{+1.0}_{-0.8}$ kpc \citep{Bailer-Jones+21}. According to \citet{Negueruela+96}, it is classified as an O8-9Ve star, while for \citet{MaizApellaniz+16} it is a star of O5Ve type; very likely, the reason for this disagreement between the two spectral classifications is that only \citet{Negueruela+96} corrected for the He infilling emission. Subsequent \RXTE\ observations led to the discovery of pulsed emission ($P_{\rm spin}$ = 103.2$\pm$0.1 s) and to the detection of the orbital period ($P_{\rm orb}$ $\simeq$ 34.5 days; \citealt{CorbetPeele97}). The source monitoring performed between 2005 and 2016 with the BAT instrument of the \Swift\ \textit{Neil Gehrels Observatory} confirmed the orbital period and revealed an almost steady emission with no brightening \citep{Corbet+16}. The source was also included in the catalogues of persistent X-ray sources detected with \SAX\ WFC \citep{Capitanio+11} and INTEGRAL IBIS \citep{Bird+16}. The latest observation of \four\ was performed in 2016 with \Astro, when the source was in a non-flaring persistent state at a flux level \fx\ (0.4-7 keV) = 6.6$\times 10^{-11}$ \flux. It provided a new measure of the pulse period ($P_{\rm spin}$ = 103.144$\pm$0.001 s) and revealed an energy-dependent pulse profile and a broad (with width $\sigma$ = 1.1 keV) iron line at $E \simeq$ 6.3 keV in the energy spectrum \citep{Roy+20}.

The almost constant X-ray flux of \four\ over long timescales, and its long orbital and spin periods, strongly suggest that also this source belongs to the class of the X-Persei--like persistent BeXRBs. On the other hand, the observed X-ray flux and estimated distance imply a rather high luminosity (\lx $\sim$ a few $10^{35}$ \lum) compared with the other sources of this type. Therefore, we observed it with \XMM\ to investigate its timing and spectral properties.

	\section{Observation and data reduction}
	\label{data}

The \XMM\ observation of \four\ was performed on April 11, 2023 (MJD 60045) and its duration was $\simeq$ 33.5 ks. The source was observed with both the \pn\ \citep{Struder+01} and the two \MOS\ \citep{Turner+01} cameras of the \EPIC\ instrument; in addition, also the Reflection Grating Spectrometer (\RGS) was used \citep{denHerder+01}. Due to the rather high source flux, which implies a high count rate (CR), for the \EPIC\ cameras the \textit{Small Window} operating mode was used, in order to prevent photon pileup. The configuration of the observing instruments is reported in Table~\ref{observation}, where the net exposure time of the \EPIC\ cameras does not include the readout dead time, which contributes for 2.9 \% and 29.9 \% for the \MOS\ and \pn\ cameras, respectively.

We used the version 21 of the Science Analysis System (\texttt{SAS}) to process the collected data. First of all, we looked for the possible presence of periods characterized by a high rate of soft protons (SP), which for \EPIC\ could affect the results of the data analysis. We verified that no SP flares occured during the observation: the rate of SPs was negligible along the whole exposure and, then, we used the whole exposure time.

In the case of the \EPIC\ data, we considered events with pattern in the range 0-12 (corresponding to up to 4 contiguous pixels) for the \MOS\ data and in the range 0-4 (i.e. mono- and bi-pixels) for the \pn\ data. In all \EPIC\ cameras the source position was very near the window edge. Therefore, we were forced to select source events only from a circular region, centered at the source position, with a small radius of 30 arcsec. In the case of the \pn\ camera we had the same constraint also for the selection of the background events, for which we considered a circular extraction region of the same size; instead for both the \MOS\ cameras we considered a background extraction radius of 120 arcsec. Finally, we verified also that the data were not affected by photon pileup: in fact, although the count rate was rather high (Table~\ref{observation}), it was well below the threshold for significant pile-up estimated by \citet{Jethwa+15}.

Our analysis revealed that the \RGS\ data do not show the presence of any significant emission or absorption feature and provided no further information compared with the \EPIC\ ones. Therefore, in the following we will not discuss them.

\begin{table*}
\caption{Summary of the \XMM\ observation of \four\ (ID 0901280101).}\label{observation}
\vspace{-1 cm}
\begin{center}
\begin{tabular}{ccccccc} \\ \hline
Instrument      & Filter        & Mode                  & Time Resolution       & Net Exposure Time       & Extraction Radius     & Net Count Rate  \\
                &               &                       &                       & (ks)                    & (arcsec)              & (\countsec)     \\ \hline
\pn\            & Thin 1        & Small Window          & 5.7 ms                & 23.2                    & 30                    & 2.8$\pm$1.6 \\
MOS1            & Medium        & Small Window          & 0.3 s                 & 32.6                    & 30                    &  
1.0$\pm$0.6  \\
MOS2            & Medium        & Small Window          & 0.3 s                 & 32.6                    & 30                    &  
1.0$\pm$0.6  \\
RGS1            & -             & Spectroscopy          & 4.8 s           
      & 33.9                    & -                     & - \\
RGS2            & -             & Spectroscopy          & 9.6 s           
      & 33.8                    & -                     & - \\ \hline
\end{tabular}
\end{center}
\end{table*}

	\section{Timing analysis}
	\label{timing}

\begin{figure}
\begin{center}
\includegraphics[height=8.5cm,angle=-90]{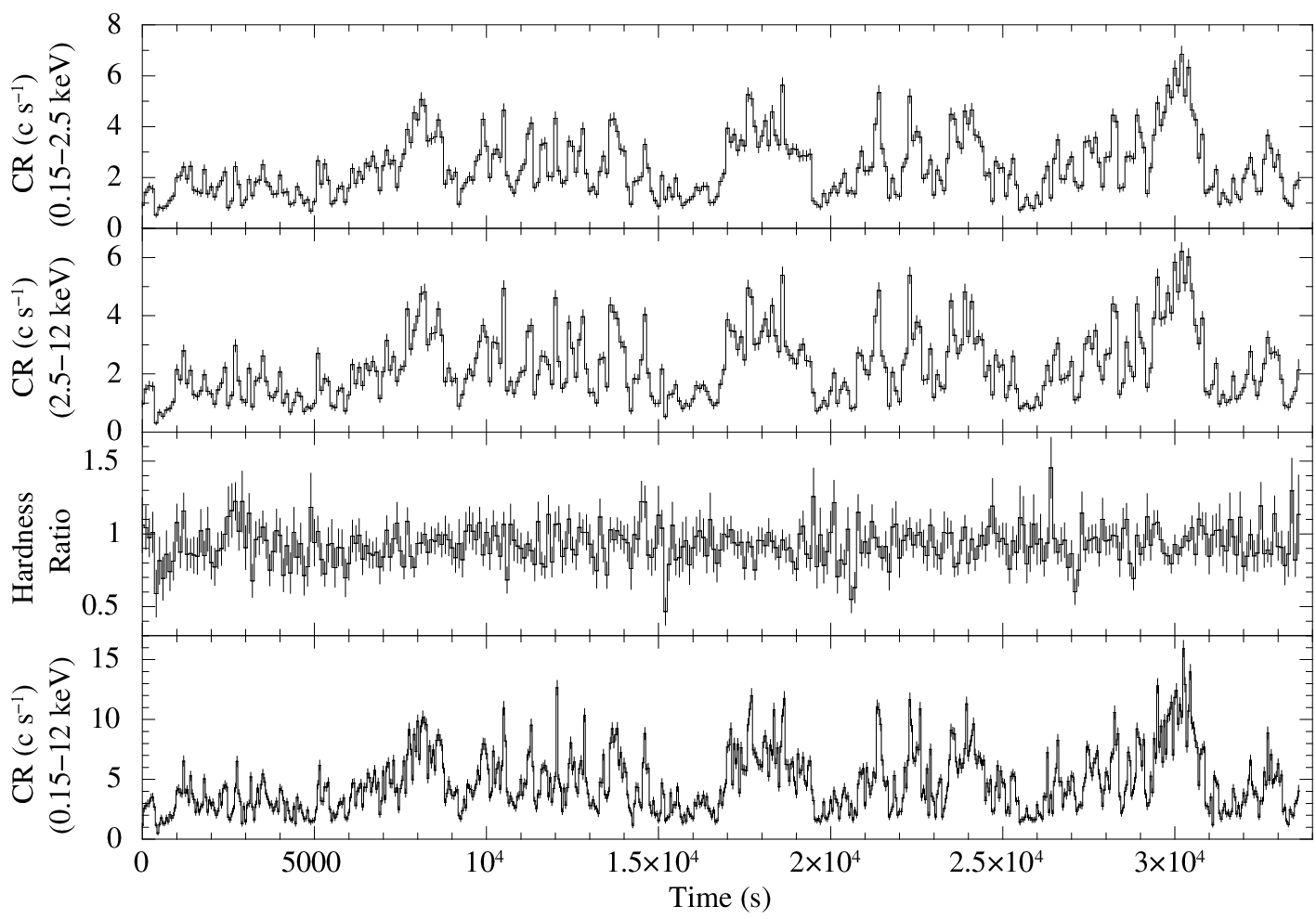}
\caption{{\small Light curves of \four, corrected for the background and the extraction region, in the \textit{S} (top panel) and \textit{H} (second panel) energy ranges, together with their ratio (third panel); the binning time is 100 s/bin. In the bottom panel the curve in the full energy range is reported, with a binning time of 50 s/bin.}}
\label{lc}
\end{center}
\vspace{-0.75 cm}
\end{figure}

The arrival times of the \EPIC\ events were reported to the Solar System barycentre by using the \texttt{SAS} tool \textsc{barycenter}. In our timing analysis, we considered both the full energy range 0.15--12 keV and two sub-ranges, one \textit{soft} (\textit{S}) between 0.15 and 2.5 keV and one \textit{hard} (\textit{H}) between 2.5 and 12 keV: in this way, we obtained a similar source count number in both ranges. For each of the three cameras, we accumulated a light curve in each of these three energy ranges; then, we used the \texttt{SAS} tool \textsc{epiclccorr} to properly take into account both the background and the extraction region. In this way we obtained a net source CR in the full energy range of $\simeq$ 2.8 \countsec\ and $\simeq$ 1.0 \countsec\ for, respectively, the \pn\ camera and each of the two \MOS\ cameras. Afterwards, for each energy range we summed the light curves of the three cameras to obtain the overall \EPIC\ light curves. These curves are shown in Fig.~\ref{lc}, toghether with the hardness ratio (HR) between the \textit{H} and the \textit{S} light curves (HR = \textit{H}/\textit{S}). The average CR is $\simeq$ 2.5 and $\simeq$ 2.3 \countsec\ for the \textit{S} and the \textit{H} range, respectively.

The light curves reported in Fig.~\ref{lc} clearly show that \four\ is characterized by a very large flux variability on short time scales: in both the partial ranges, where the bin time is 100 s, the CR varies by a factor up to $\sim$ 3 between two consecutive bins. This is confirmed by the large values of the CR standard deviation reported in Table~\ref{observation}. This finding is further confirmed in the full energy range, where the binning time is 50 s only: also in this case, indeed, we can observe a comparable variability factor between two nearby bins. On the other hand, the source is rather stable on the long time scales, since all the large short-scale variations occur always around the same average CR value. Even the HR is rather constant: it shows variations of $\sim$ 50 \% at most and it is not correlated with the source CR.

\begin{figure}
\begin{center}
\includegraphics[height=8.5cm,angle=-90]{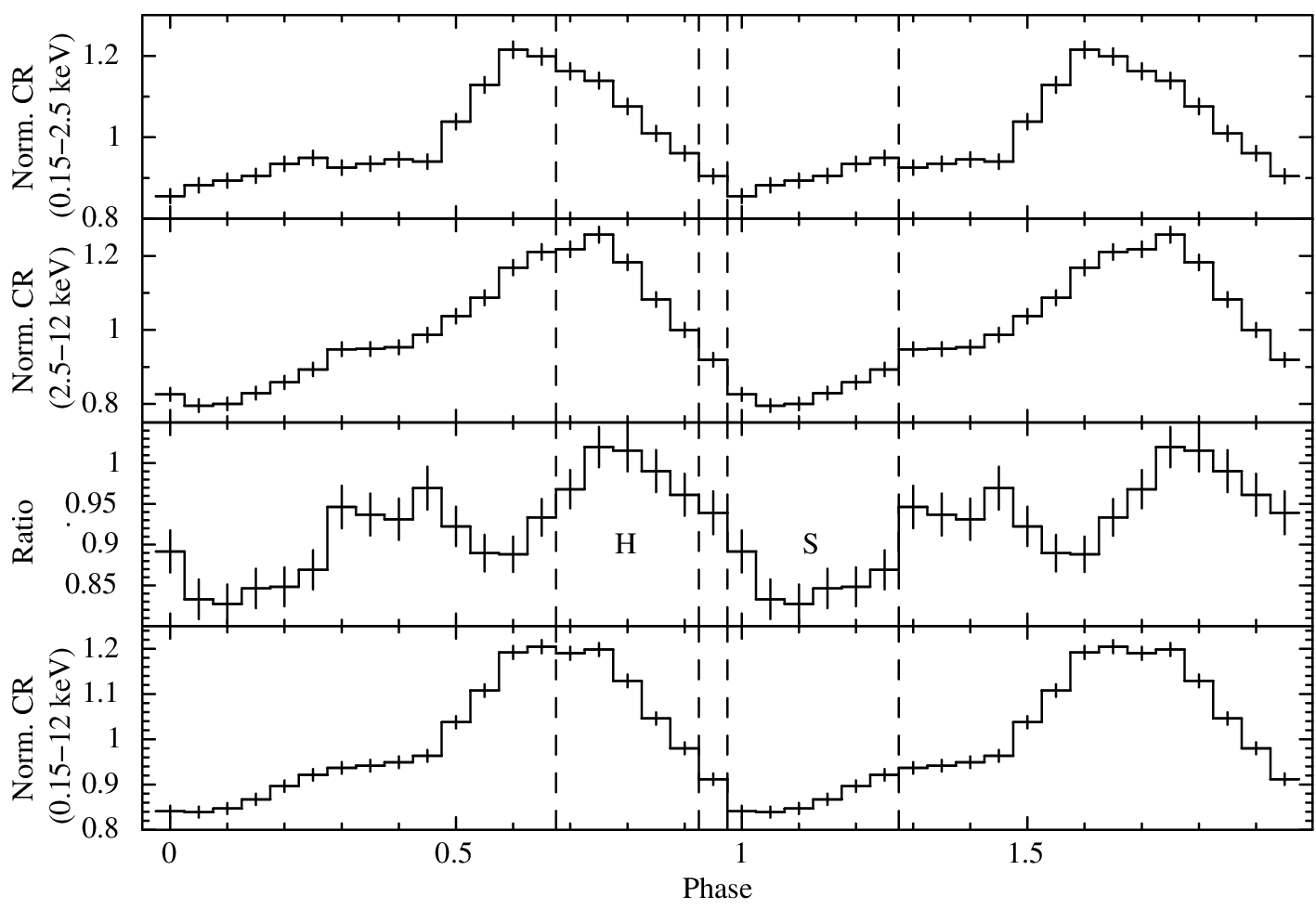}
\caption{{\small Light curves of \four\ in the full (bottom panel), \textit{S} (top panel) and \textit{H} (second panel) energy ranges, together with the HR (third panel), folded at the pulse period $P$ = 103.301 s. The phase ranges selected for the phase-resolved spectral analysis are delimited with vertical dashed lines.}}
\label{flc2E}
\end{center}
\vspace{-0.75 cm}
\end{figure}

Once corrected for the Solar System barycentre, the datasets of the three instruments were merged into a unique event list to increase the count statistics. The best-fit period, obtained with the phase-fitting technique, is $P$ = 103.301(5) s. Based on this period, we built the folded light curve in each of the three energy ranges, together with the folded HR between the \textit{H} and the \textit{S} curves. They are reported in Fig.~\ref{flc2E}. In both energy ranges, the pulse profile is characterized by a single broad peak, which extends for almost half of the pulse period. Both the rise and the decay of the CR are rather smooth, but in the \textit{H} range the peak is slightly delayed compared with the \textit{S} range. The HR indicates some spectral variation during the rising part of the pulse.

\begin{figure}
\begin{center}
\includegraphics[height=8.5cm,angle=-90]{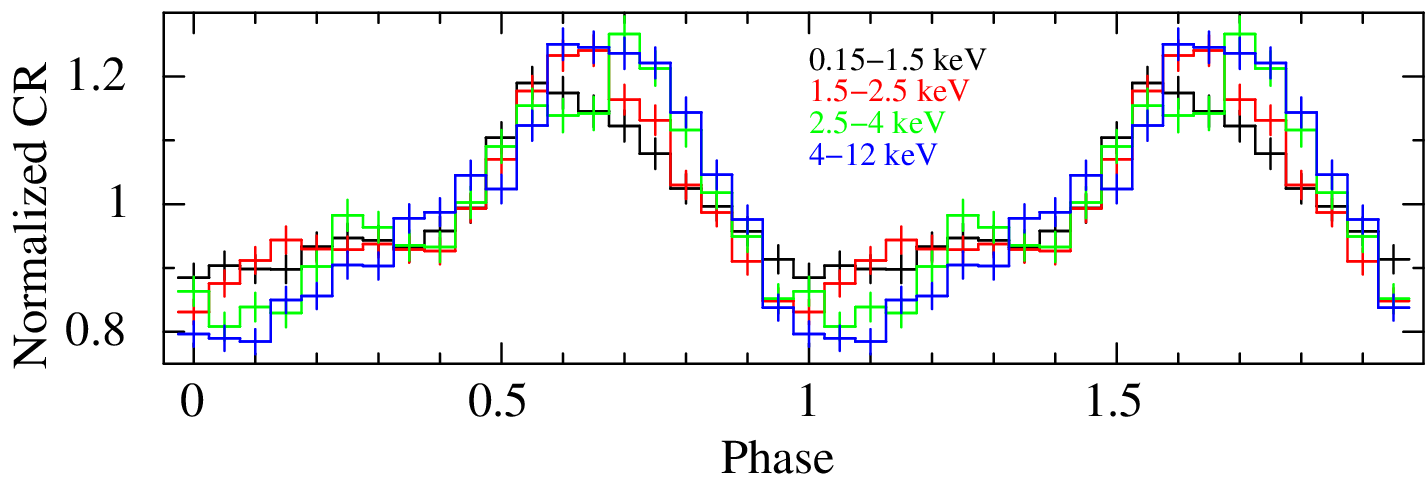}
\caption{{\small Pulse profile of \four\ in the energy ranges 0.15--1.5, 1.5--2.5, 2.5--4, and 4--12 keV.}}
\label{flc4E}
\end{center}
\vspace{-0.75 cm}
\end{figure}

The dependence of the pulse profile on the photon energy is confirmed by Fig.~\ref{flc4E}, where we report the folded light curves in four narrower energy ranges: 0.15-1.5, 1.5-2.5, 2.5-4, and 4-12 keV. As in the previous case, all of them show a single broad peak, with no specific features. The flux variability shows a clear dependence with the photon energy: the average pulsed fraction (PF), defined as PF = (CR$_{\rm max}$ - CR$_{\rm min}$)/(2$\times$CR$_{\rm average}$), gradually increases from $\simeq$ 15 \% at E $<$ 1.5 keV to $\simeq$ 23 \% at E $>$ 4 keV.

	\section{\EPIC\ time-averaged spectral analysis}
	\label{average-spectroscopy}

Although the source flux of \four\ was highly variable on short timescales during the \XMM\ observation, over the whole observation it showed neither an increasing nor a decreasing trend. In addition, the HR showed a limited variability (Fig.~\ref{lc}), thus indicating that the spectral properties were almost constant. For these reasons we performed the spectral analysis of the source over the whole exposure, with no time selection. To this aim, we obtained a unique source spectrum for each \EPIC\ camera by selecting events from the same extraction regions already considered for the light curves. We used the \texttt{SAS} tool \textsc{specgroup} to perform the rebinning of each spectrum with a minimum significance of 3~$\sigma$ for each energy bin. Then, the spectral channels consistent with zero (after the background subtraction) were removed. The response matrices and ancillary files were produced using, respectively, the \texttt{SAS} tasks \textsc{rmfgen} and \textsc{arfgen}. The spectral analysis was performed using version 12.12.1 of \texttt{XSPEC} \citep{Arnaud96}; to this aim we considered the energy range between 0.2 and 12 keV. In all cases the spectral uncertainties were calculated at the 90 \% confidence level for each interesting parameter and we considered a source distance $d$ = 7.6 kpc. The spectral fitting was performed on the three \EPIC\ spectra simultaneously, since there was no evidence that the separate fits provided inconsistent results. This was done by introducing a free normalization parameter among the three cameras, in order to properly take into account the possible uncertainties in the instrumental responses: more precisely, once the normalisation factor of the \pn\ spectrum was fixed to 1, we obtained relative normalisation factors 0.938 $\pm$ 0.008 and 1.091 $\pm$ 0.008 for, respectively, the \textit{MOS1} and the \textit{MOS2} spectra. In our analysis we used the elemental abundances reported by \citet{WilmsAllenMcCray00}, the photoelectric absorption cross-sections provided by \citet{Verner+96}, and the absorption model \textsc{tbabs} defined in \texttt{XSPEC}. The estimated fluxes were calculated with the \texttt{XSPEC} tool \textsc{cflux}.

Our first attempt was to describe the \EPIC\ continuum spectrum with a simple power-law (PL) model that provides a phenomenological description of the emission due to the matter accretion from the companion star onto the neutron star: with it the best-fit values of the photon index and interstellar absorption were, respectively, $\Gamma$ = 1.33$^{+0.2}_{-0.1}$ and $N_{\rm H}$ = (6.9$^{+0.1}_{-0.2}$) $\times 10^{21}$ cm$^{-2}$, and the corresponding reduced chi-squared was \chisqnu\ = 1.215 for 2508 degrees of freedom (d.o.f.). Starting from this model, we obtained a significantly better result in two alternative ways: either with the introduction of a partial covering fraction absorption (\textsc{tbpcf} in \texttt{XSPEC}), which describes the absorption due to an inhomogeneous absorber medium, or by replacing the PL model with a cutoff power law (CPL). In both cases the reduced chi-squared decreased to \chisqnu\ = 1.063, with 2506 and 2507 degrees of freedom (d.o.f.), respectively, and the observed source flux is \fx\ $\simeq$ 1.7$\times 10^{-11}$ \flux. We tested also models with a single thermal component, but in all cases we obtained a worse fit.

\begin{figure}
	\begin{center}
		\includegraphics[height=8.5cm,angle=-90]{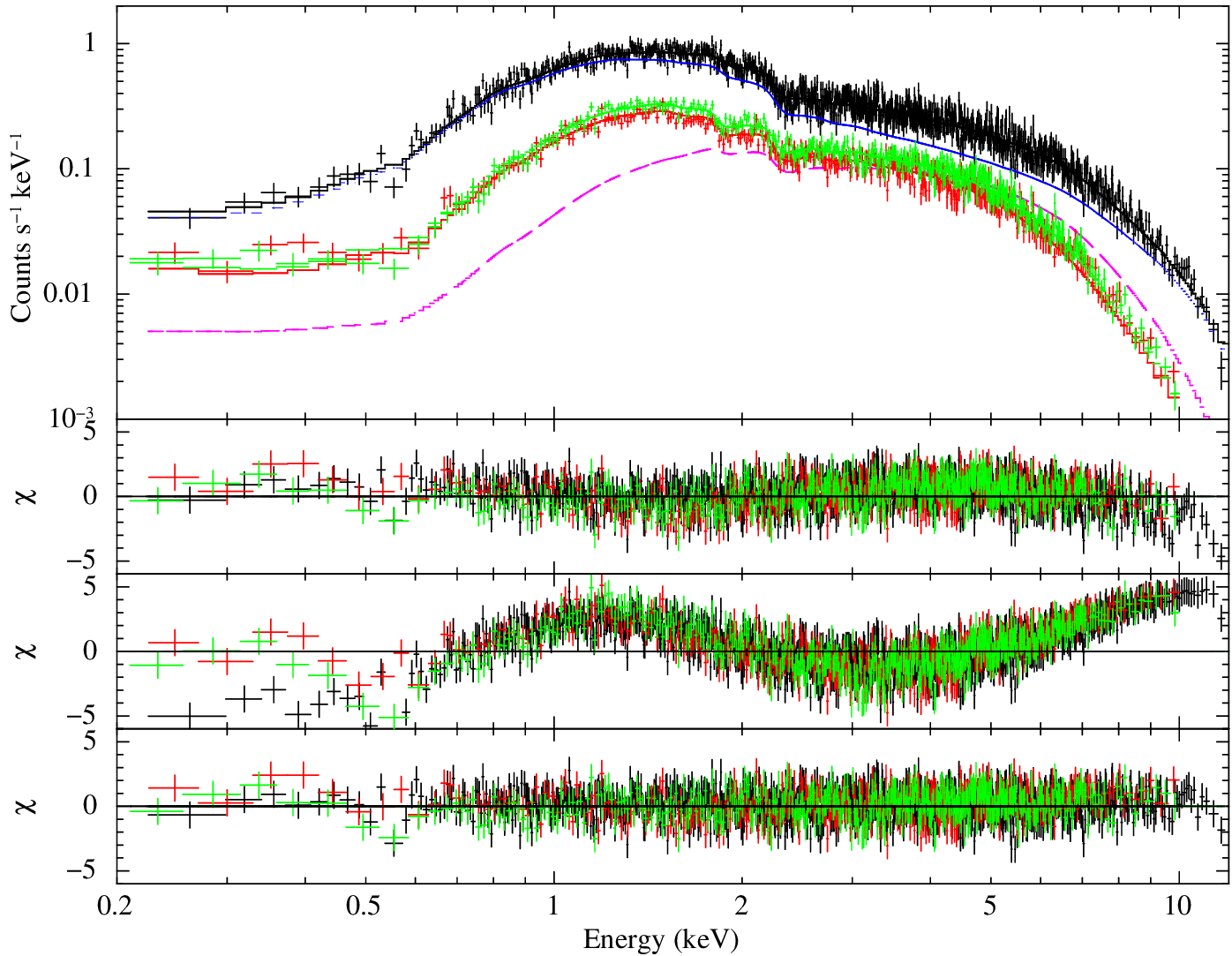}
		\caption{{\small Time-averaged spectrum of \four, where the black, red, and green symbols represent the \pn, \textit{MOS1}, and \textit{MOS2} data, respectively. \textit{Upper panel}: superposition of the best-fitting PL+BB model, for the \pn\ spectrum only, with the \textit{EPIC} spectra. \textit{Middle panels}: residuals (in units of $\sigma$) between data and model in the case of the single PL and of the single BB. \textit{Bottom panel}: residuals in the case of the PL+BB.}}
		\label{average-spectrum}
	\end{center}
	\vspace{-0.75 cm}
\end{figure}

We obtained a slightly better fit (\chisqnu/d.o.f. = 1.058/2505) with a CPL plus a component due to collisionally ionised gas (\textsc{apec} model in \texttt{XSPEC}) that accounts for the thermal emission from a possible optically thin gas around the accreting NS. An even better fit (\chisqnu/d.o.f. = 1.050/2506) is obtained with a PL plus a black body (BB), where the latter component describes the thermal emission of a possible optically thick plasma on the NS surface, a scenario that is often observed in this type of sources. In both cases the non-thermal component (either the PL or the CPL) dominates over the thermal one (either the BB or the APEC). The F-test statistics value is very high ($\simeq 200$) for the PL+BB model, which implies a very negligible probability \textbf{($< 10^{-80}$)} that the additional BB component is an artifact. This is confirmed by the spectrum reported in Fig.~\ref{average-spectrum}, where we show that the data-model residuals are much larger in the case of the fit with a PL or a BB model, compared with the PL+BB model. For the CPL+APEC model, instead, the F-test statistics value is much lower ($\simeq 7.7$) and, then, the probability that the APEC component is spurious is just below $10^{-3}$. Apart from these two models, with other two-component models it was not possible to obtain equally good fits. In particular, the PL+APEC model resulted in a significantly worse fit (\chisqnu/d.o.f. = 1.182/2506), while the best-fit CPL+BB model was fully equivalent to the PL+BB model, since E$_{\rm cut} >$ 100 keV.

We noted that the fit of the continuum spectrum with each of the previous models left weak residuals at some specific energies, such as at $\simeq$ 0.55 keV. We verified that it was possible to describe these features with a Gaussian model, but further investigations of these features with Monte Carlo simulations, performed using the \texttt{XSPEC} routine \textsc{simftes}, estimated a high probability that these features are not significant. Therefore, in the following analyses we will ignore them. Moreover, we looked for the possible presence of emission lines due to iron $K_\alpha$ transition, at energies between 6 and 7 keV and with widths in the range 0-0.5 keV (in steps of 0.1 keV). We did not find any feature of this type, with an upper limit of $\lsim$ 100 eV (at 90 \% c.l.) on its equivalent width.

The best-fitting parameters we obtained by using the four previous models are reported in Table~\ref{spectral_parameters}. All models are characterozed by a moderate interstellar absorption, since \nh $\sim (5-7) \times 10^{21}$ cm$^{-2}$, and a rather hard non-thermal component, since $\Gamma \simeq 0.7$ in the case of the CPL and 1.5--1.8 in the case of the PL; moreover, the cut-off energy of the CPL model is rather low. The unabsorbed flux of the non-thermal component is always \fx\ $\sim (2-3)\times 10^{-11}$ \flux\ and, in the cases of the two-component models, contributes most of the source flux and dominates over the thermal component at all energies. Both the additional thermal components have a rather high temperature but, while in the case of the BB it provides a significant fraction of the total source flux, in the case of the APEC its contribution is only marginal.

\begin{table*}[!t]
	\caption{Best-fit parameters of the time-averaged \textit{EPIC} spectrum of \four, in the case of the four best-fitting continuum models.}\label{spectral_parameters}
	\begin{center}
		\begin{tabular}{cccccc} \hline \hline
			Model								& -		& TBPCF$\times$PL	& CPL			& PL+BB			& CPL+APEC	\\
			Parameter							& Unit	& Value             & Value         & Value			& Value	\\ \hline
			TBABS \nh							& $\times10^{21}$  cm$^{-2}$	& 7.5$^{+0.3}_{-0.2}$	& 5.2$^{+0.1}_{-0.2}$	& 6.4$^{+0.3}_{-0.2}$		& 5.2$\pm$0.2	\\
			TBPCF \nh							& $\times10^{22}$  cm$^{-2}$	& 6.4$^{+0.6}_{-0.5}$	& -	& -	& -	\\
			TBPCF Covering Fraction				& -		& 0.51$\pm$0.03	& -	& -	& -	\\
			$\Gamma$							& -		& 1.79$^{+0.05}_{-0.04}$	& 0.73$^{+0.05}_{-0.06}$		& 1.52$\pm$0.06		& 0.66$\pm$0.06	\\
			E$_{\rm cut}$						& keV	& -		& 6.5 $^{+0.7}_{-0.5}$		& -			& 6.0$^{+0.6}_{-0.4}$	\\
			Flux$_{\rm PL or CPL}$ (0.2-12 keV)$^{(a)}$	& $\times 10^{-11}$ \flux	& 3.3$\pm$0.2	& 1.91$\pm$0.03	& 1.57$^{+0.06}_{-0.07}$	& 1.88$\pm$0.03	\\
			$kT_{\rm BB~or~APEC}$				& keV	& -		& -			& 1.48$^{+0.07}_{-0.07}$		& 1.4$^{+0.4}_{-0.3}$	\\
			$R_{\rm BB}^{(b)}$					& m		& -		& -			& 240$^{+30}_{-20}$	& -	\\
			$N_{\rm APEC}$						& cm$^{-5}$	& -	& -			& -			& 1.4$^{+1.3}_{-0.8} \times 10^{-4}$	\\
			Flux$_{\rm BB~or~APEC}$ (0.2-12 keV)$^{(a)}$	& $\times 10^{-12}$ \flux	& -	& -			& 5.1$\pm$0.7	& 0.2$^{+0.2}_{-0.1}$	\\ \hline
			Flux$_{\rm BB~or~APEC}$/Flux$_{\rm TOT}$ (0.2-12 keV)	& -	& -	& -			& 24.2 \%	& 1.1 \%	\\
			Flux$_{\rm BB~or~APEC}$/Flux$_{\rm TOT}$ (0.01-100 keV)	& -	& -	& -			& 9.5 \%	& 1.1 \%	\\
			Unabsorbed flux (0.2-12 keV)		& $\times 10^{-11}$ \flux	& 3.3$\pm$0.2		& 1.91$\pm$0.03	& 2.07$^{+0.03}_{-0.02}$	& 1.90$^{+0.03}_{-0.02}$	\\
			Luminosity (0.2-12 keV)$^{(b)}$		& $\times 10^{35}$ \lum 	& 2.2$\pm$0.1		& 1.25$\pm$0.02	& 1.35$^{+0.02}_{-0.01}$	& 1.24$^{+0.02}_{-0.01}$	\\
			\chisqnu/d.o.f.						& -							& 1.063/2506		& 1.063/2507	& 1.050/2506			& 1.058/2505	\\ \hline
		\end{tabular}
	\end{center}
	Notes: $^{(a)}$ Corrected for absorption; $^{(b)}$ Assuming a source distance $d$ = 7.6 kpc
\end{table*}

	\section{\EPIC\ phase-resolved spectral analysis}
	\label{resolved-spectroscopy}

Fig.~\ref{flc2E} reveals that the HR is variable along the pulse phase, with a primary and a secondary peak just after and before the flux maximum, respectively. Since this implies that the source spectrum varies depending on the spin phase, we performed a specific spectral analysis in two distinct phase ranges: for each \textit{EPIC} camera, we accumulated a \textit{hard} spectrum (H) in the phase range $\Delta\phi$ = 0.675-0.925 (where HR $>$ 0.95) and a \textit{soft} spectrum (S) in the phase range $\Delta\phi$ = 0.975-1.275 (where HR $<$ 0.9). In this way, the two spectra are characterized by very similar number of source counts, in spite of the different length of the two phase intervals (0.25 and 0.3 for, respectively, spectrum H and S).

Our first aim was to verify if the best-fitting models of the time-averaged spectrum can provide a satisfactory description also of the two phase-resolved spectra. Therefore, we performed an independent fit of the two spectra using the four models described in Section~\ref{average-spectroscopy}. The results obtained with this approach are reported in Table~\ref{phase-resolved-parameters}. We found that both spectra can be described with each model and that in all cases the source flux decreases from (1.88$\pm$0.05) $\times 10^{-11}$ \flux\ for spectrum H to (1.34$\pm$0.04)$\times 10^{-11}$ \flux\ for spectrum S, corresponding to a reduction of $\simeq$ 28 \%. However, for each model there is a significant difference of the best-fit parameter values between the two spectra: this holds for the interstellar absorption, the photon index of the non-thermal component and, in the case of the PL+BB model, also for the temperature of the thermal component. In the case of the PL+BB model the relative contribution of the two components to the total flux does not vary. For the CPL+APEC model, instead, it is not possible to make any comparison, due to the large uncertainties in the parameters of the APEC component.

\begin{table}[!t]
	\caption{Best-fit parameters of the \textit{EPIC} spectra H and S of \four, in the case of the four best-fitting continuum models.}\label{phase-resolved-parameters}
	\begin{center}
		\begin{tabular}{ccc} \hline \hline
			Parameter					            & Spectrum H                & Spectrum S			\\ \hline
			\multicolumn{3}{c}{TBPCF$\times$PL}    \\
			TBABS \nh ($\times10^{21}$ cm$^{-2}$)	& 7.3$\pm$0.4				& 7.8$\pm$0.4			\\
			TBPCF \nh ($\times10^{22}$ cm$^{-2}$)	& 6.3$\pm$1.1				& 6.8$^{+1.4}_{-1.3}$	\\
			TBPCF Covering Fraction	                & 0.51$^{+0.05}_{-0.06}$	& 0.46$^{+0.06}_{-0.07}$	\\
			$\Gamma$					            & 1.70$^{+0.09}_{-0.08}$	& 1.83$^{+0.09}_{-0.10}$   \\
			Flux$_{\rm PL}$ (0.2-12 keV)$^{(a)}$	& 3.4$^{+0.8}_{-0.3}$	    & 2.7$^{+0.4}_{-0.3}$	\\
			Luminosity (0.2-12 keV)$^{(b)}$			& 2.2$^{+0.3}_{-0.2}$	    & 1.8$\pm$0.2			\\
			\chisqnu/d.o.f.				            & 1.047/1615                & 0.974/1527            \\ \hline
			\multicolumn{3}{c}{CUTOFFPL}    \\
			TBABS \nh ($\times10^{21}$ cm$^{-2}$)	& 4.9$^{+0.4}_{-0.3}$	    & 5.8$^{+0.3}_{-0.4}$	\\
			$\Gamma$					            & 0.65$^{+0.11}_{-0.10}$	& 1.0$\pm$0.1			\\
			E$_{\rm cut}$ (keV)	                    & 6.6$^{+1.4}_{-0.9}$       & 9$\pm$2				\\
			Flux$_{\rm CPL}$ (0.2-12 keV)$^{(a)}$	& 2.10$\pm$0.06				& 1.61$\pm$0.05	        \\
			Luminosity (0.2-12 keV)$^{(b)}$			& 1.37$\pm$0.04             & 1.05$\pm$0.03			\\
			\chisqnu/d.o.f.				            & 1.048/1616                & 0.980/1528			\\ \hline
			\multicolumn{3}{c}{PL+BB}    \\
			TBABS \nh ($\times10^{21}$ cm$^{-2}$)	& 5.9$\pm$0.4				& 6.8$\pm$0.4	        \\
			$\Gamma$					            & 1.33$^{+0.11}_{-0.08}$	& 1.6$\pm$0.1			\\
			Flux$_{\rm PL}$ (0.2-12 keV)$^{(a)}$	& 1.8$^{+0.1}_{-0.2}$		& 1.41$^{+0.10}_{-0.09}$\\
			$kT_{\rm BB}$ (keV)	                    & 1.4$^{+0.1}_{-0.2}$		& 1.5$^{+0.1}_{-0.2}$	\\
			$R_{\rm BB}$ (m)$^{(c)}$               	& 270$\pm$40				& 200$\pm$30			\\
			Flux$_{\rm BB}$ (0.2-12 keV)$^{(d)}$	& 5$\pm$1					& 3$\pm$1				\\
			Flux$_{\rm BB}$/Flux$_{\rm TOT}$ (0.2-12 keV)	& 20.6 \%	        & 19.0 \%	            \\
			Flux$_{\rm BB}$/Flux$_{\rm TOT}$ (0.01-100 keV)	& 5.7 \%	        & 7.6 \%	            \\
			Unabsorbed flux (0.2-12 keV)$^{(a)}$	& 2.23$\pm$0.05				& 1.74$^{+0.07}_{-0.05}$\\
			Luminosity (0.2-12 keV)$^{(b)}$			& 1.46$\pm$0.03				& 1.14$^{+0.05}_{-0.03}$\\
			\chisqnu/d.o.f.				            & 1.044/1615				& 0.975/1527			\\ \hline
			\multicolumn{3}{c}{CPL+APEC}    \\
			TBABS \nh ($\times10^{22}$ cm$^{-2}$)	& 4.9$^{+0.4}_{-0.3}$		& 5.8$^{+0.8}_{-0.4}$   \\
			$\Gamma$					            & 0.6$\pm$0.1				& 0.9$\pm$0.1           \\
			E$_{\rm cut}$ (keV)	                    & 6$^{+2}_{-1}$				& 8$\pm$2				\\
			Flux$_{\rm CPL}$ (0.2-12 keV)$^{(a)}$	& 2.06$^{+0.08}_{-1.4}$		& 1.58$^{+0.07}_{-0.13}$\\
			$kT_{\rm APEC}$ (keV)	                & 2.2$\pm$1.7				& 1.4$^{+1.5}_{-1.0}$	\\
			$N_{\rm APEC}$ (cm$^{-5}$)				& $< 8\times 10^{-4}$		& 1.2$^{+3.1}_{-1.1}$	\\
			Flux$_{\rm APEC}$ (0.2-12 keV)$^{(e)}$	& -							& 1.8$^{+4.4}_{-0.9}$	\\
			Flux$_{\rm APEC}$/Flux$_{\rm TOT}$ (0.2-12 keV)     & -				& 1.1 \%	            \\
			Flux$_{\rm APEC}$/Flux$_{\rm TOT}$ (0.01-100 keV)	& -				& 1.1 \%	            \\
			Unabsorbed flux (0.2-12 keV)$^{(a)}$	& 2.09$^{+0.06}_{-0.05}$	& 1.60$^{+0.07}_{-0.05}$\\
			Luminosity (0.5-10 keV)$^{(b)}$			& 1.36$^{+0.04}_{-0.03}$	& 1.04$^{+0.05}_{-0.03}$\\
			\chisqnu/d.o.f.				            & 1.048/1614				& 0.979/1526            \\ \hline
		\end{tabular}
	\end{center}
	Notes: $^{(a)}$ Corrected for absorption, $\times 10^{-11}$ \flux; $^{(b)}$ $\times 10^{35}$ \lum, assuming a source distance $d$ = 7.6 kpc; $^{(c)}$ Assuming a source distance $d$ = 7.6 kpc; $^{(d)}$ Corrected for absorption, $\times 10^{-12}$ \flux; $^{(e)}$ Corrected for absorption, $\times 10^{-13}$ \flux
\end{table}

The previous results show that the spectral shape varies along the pulse phase. In the following, we try to investigate whether, in the PL+BB model, the variation can be ascribed to only one of the two components. In this case, we found that it was not possible to describe the two spectra by allowing a variation of only the PL flux (Null hypothesis probability $\simeq$ 0.02), while we obtained a good fit with a common PL component and variable BB flux (Null hypothesis probability $\simeq$ 0.22). We also investigated two alternative scenarios regarding the variability of the two components:
\begin{itemize}
	\item On the one hand, we forced common values between the two spectra for \nh, $\Gamma$, and $kT_{\rm BB}$, allowing only the normalization of the two components to vary (Case 1): this implies that the spectral variablity along the pulse phase is attributed to a difference in the relative contribution of the two components.
	\item On the other hand, we forced a common BB component by allowing only the PL parameters to vary (Case 2): in this case the spectral variability is fully attributed to a change in the non-thermal component.
\end{itemize}
The results obtained with this approach are reported in Table~\ref{2spectra_common}. They show that the two cases are fully comparable from the point of view of the fit quality, since they have almost equal \chisqnu: in Case 1 the flux variability is almost fully due to a variation of the BB flux, since the PL flux is comparable in the two spectra; in Case 2 the constant BB component needs that not only the flux but also the photon index of the PL component is significantly variable between the two spectra.

\begin{table}[htbp]
	\caption{Best--fit values for the BB and PL parameters, when the two spectra H and S are fitted simultaneously with common values of $N_{\rm H}$, $\Gamma$, and $kT_{\rm BB}$ (Case 1) or with common values of $N_{\rm H}$, $kT_{\rm BB}$, and $R_{\rm BB}$ (Case 2).}\label{2spectra_common}
	\begin{center}
	\begin{tabular}{c|cc|cc} \hline
		Phase	 			&		\multicolumn{2}{c}{Case 1}					&		\multicolumn{2}{c}{Case 2}				\\
		interval 			&		H 							&		S						&	H				&	S			\\ \hline
		$N_{\rm H}^{a}$		& \textit{6.7$^{+0.2}_{-0.3}$}		& \textit{6.7$^{+0.2}_{-0.3}$}	& \textit{6.4 $\pm$ 0.3}		& \textit{6.4 $\pm$ 0.3}	\\ \hline
		$\Gamma$			& \textit{1.55$^{+0.08}_{-0.07}$}	& \textit{1.55$^{+0.08}_{-0.07}$}	& 1.34$^{+0.06}_{-0.05}$		& 1.55$\pm$0.08	\\
		$f_{\rm PL}^{b}$	& 1.58$^{+0.09}_{-0.08}$			& 1.45$\pm$0.08					& 1.92$\pm$0.08		& 1.34$\pm$0.07	\\ \hline
		$kT_{\rm BB}$		& \textit{1.56 $\pm$ 0.09}			& \textit{1.56 $\pm$ 0.09}		& \textit{1.4 $\pm$ 0.1}		& \textit{1.4 $\pm$ 0.1}	\\
		$R_{\rm BB}$		& 250$\pm$20						& 170$\pm$20					& \textit{230$^{+20}_{-30}$}		& \textit{230$^{+20}_{-30}$}	\\
		$f_{\rm BB}^{b}$	& 0.68$\pm$0.09						& 0.31$\pm$0.08					& \textit{0.36$^{+0.08}_{-0.07}$}	& \textit{0.36$^{+0.08}_{-0.07}$}	\\ \hline
		$f_{\rm TOT}^{b}$	& 2.26$\pm$0.05						& 1.76$\pm$0.03					& 2.28$^{+0.05}_{-0.04}$		& 1.70$\pm$0.04	\\
		$f_{\rm PL}$/$f_{\rm TOT}$					& 69.9 \%						& 82.4 \%					& 84.2 \%				& 78.8 \%	\\
		$f_{\rm BB}$/$f_{\rm TOT}$					& 30.1 \%						& 17.6 \%					& 15.8 \%				& 21.1 \%	\\ \hline
		$\chi^{2}_{\nu}$	& 	\multicolumn{2}{c}{1.015}							& 	\multicolumn{2}{c}{1.014}						\\
		d.o.f.				& 	\multicolumn{2}{c}{3147}							& 	\multicolumn{2}{c}{3147}						\\
		NHP$^{c}$			&	\multicolumn{2}{c}{0.28}							& 	\multicolumn{2}{c}{0.29}						\\ \hline
	\end{tabular}
	\end{center}
	Notes: $^{a}$ $\times10^{21}$ cm$^{-2}$; $^{b}$ Unabsorbed flux in the energy range 0.2--12 keV, in units of $10^{-11}$ \lum; $^{c}$ Null hypothesis probability
\end{table}

	\section{Discussion}
	\label{discussion}

\begin{figure}
	\begin{center}
		\includegraphics[height=8.5cm,angle=-90]{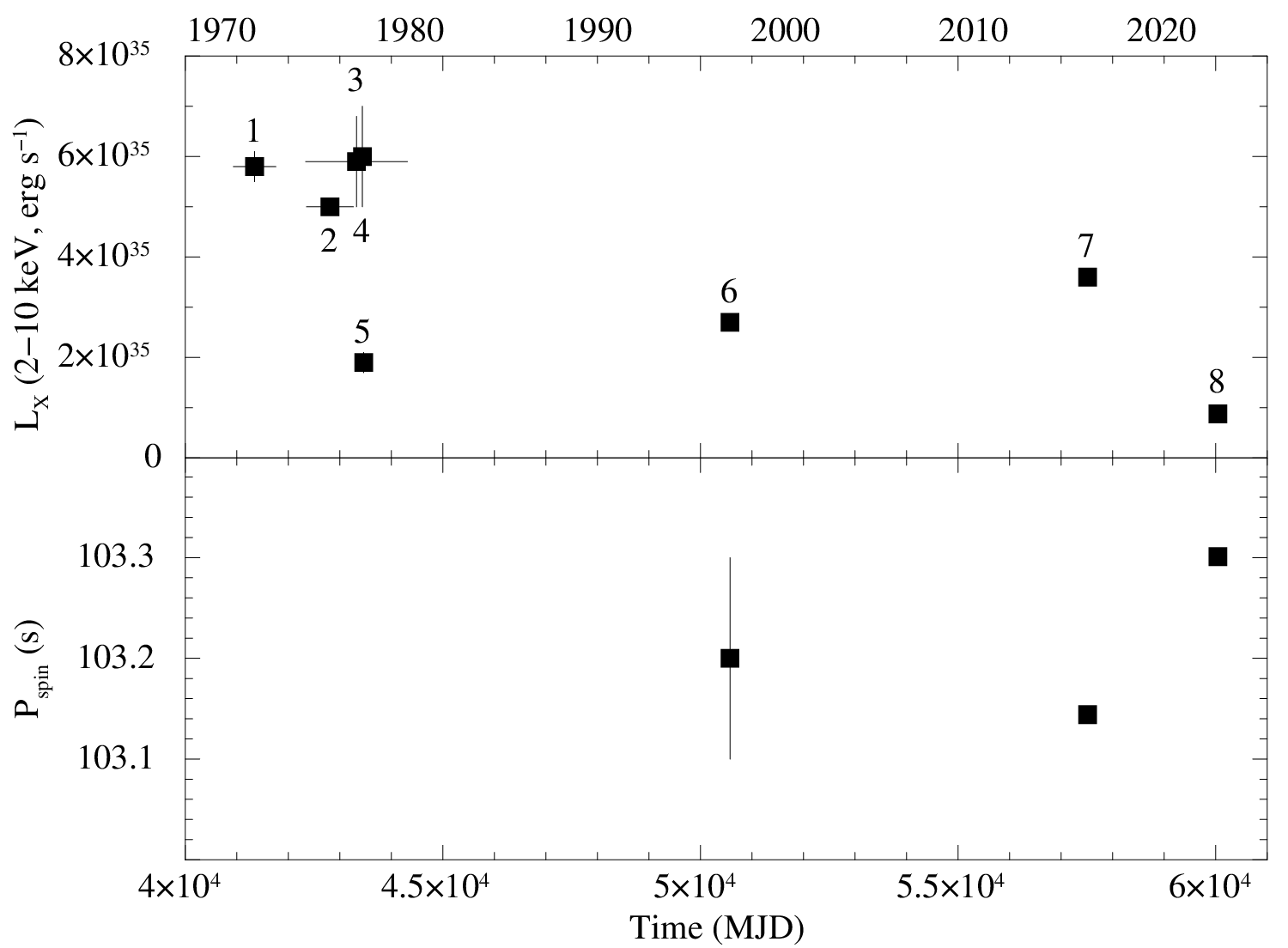}
		\caption{{\small Long-term evolution of the luminosity in the energy range 0.2-12 keV (\textit{upper panel}) and of the pulse period (\textit{lower panel}) of \four. Telescope and epoch: (1) \textit{Uhuru} catalogue: 12/12/1970 -- 18/03/1973 \citep{Forman+78}; (2) \textit{Ariel V}: 01/11/1974 -- 30/04/1977 \citep{Steiner+84}; (3) \textit{Ariel V} 3A catalogue: 15/10/1974 -- 14/03/1980 \citep{Warwick+81}; (4) \textit{HEAO1}: 19--23/10/1977 \citep{Steiner+84}; (5) \textit{HEAO1}: 15/08/1977 -- 15/02/1978 \citep{Wood+84}; (6) \textit{RXTE}: 05/05/1997 \citep{CorbetPeele97}; (7) \Astro: 06--07/05/2016 \citep{Roy+20}; (8) \XMM: 11/04/2023 (this work)}}
		\label{evolution}
	\end{center}
	\vspace{-0.75 cm}
\end{figure}

After its discovery in the '70s, \four\ was detected several times by various X-ray telescopes. In the upper panel of Fig.~\ref{evolution} we provide an overview of the X-ray luminosity levels (in the energy range 2-10 keV), which are estimated based on the flux detected by each telescope and assuming a source distance of 7.6 kpc. The first three values are based on the flux value reported in overall catalogues of X-ray sources detected with \textit{Uhuru} \citep{Forman+78} and \textit{Ariel} \citep{Warwick+81,Steiner+84}: since these catalogues were obtained with observations performed over a long time span, the reported luminosity values should be considered as average values (as indicated by the horizontal error bars). For these observations, and also for the detections obtained with \textit{HEAO-1} \citep{Steiner+84,Wood+84}, we estimated the 2-10 keV flux based on the published CR values and the appropriate count-rate-to-flux conversion factors, assuming a Crab-like PL spectrum; instead, in the case of the \RXTE\ and \Astro\ observations, we used the reported best-fit models to estimate the source flux. The figure shows that the first four detections revealed \four\ at a luminosity level in the range (5-6)$\times 10^{35}$ \lum, while in the observations 5-8 the luminosity level remained always in the range (1-4)$\times 10^{35}$ \lum; moreover, the last observation performed with \XMM\ in 2023 revealed \four\ at the lowest flux ever, with \lx\ $\simeq 9 \times 10^{34}$ \lum. The reported values show that, over a time interval of about 50 years, \four\ was characterized by a remarkable stability, since its flux varied less than a factor 10; the corresponding luminosities were always in the range $\sim 10^{35-36}$ \lum.

The pulse period of 103.301$\pm$0.005 s measured with \XMM\ in 2023 is $\simeq$ 0.15 \% longer than that of 103.144$\pm$0.001 s measured with \Astro\ in 2016 (Fig~\ref{evolution}, lower panel) and cannot be attributed to the NS orbit around the companion Be star. In fact, assuming a mass $M_{\rm Be} \sim 10 \msun$, from the orbital period $P_{\rm orb} \simeq$ 34.5 days we can estimate an average orbital velocity $v_{\rm orb} \simeq$ 140 km s$^{-1}$, which implies a variation of the spin period $\Delta P_{\rm spin} = P_{\rm spin} \times v_{\rm orb}/c \simeq$ 0.05 s. Since this value is about three times lower than the $dP_{\rm spin}$ between the \XMM\ and \Astro\ observations, the observed $P_{\rm spin}$ increase is most likely due to the NS spin down, which occured at an average rate $\pdot = 7.2 \times 10^{-10}$ s s$^{-1}$ during the seven years which separates the two observations. For comparison with other persistent BeXRBs, we note that this value is a factor $\sim$ 10 lower than that estimated for RX J0440.9+4431 during its quiescence phase \citep{LaPalombara+12} and very similar to that measured for CXOU J225355.1+624336 \citep{LaPalombara+21}. The increase in the spin period suggests that the matter transfer from the Be star onto the NS occurs via wind accretion, without any formation of an accretion disk.

The pulse shape of \four\ observed with \XMM\ is characterized by a single broad peak at all energies, which extends for almost half of the pulse period.
It is very similar to that revealed with \RXTE\ and \Astro, which also observed a single-peak profile; moreover, the same type of pulse shape is typical of several other persistent BeXRBs, as in the case of RX J0146.9+6121 \citep{LaPalombara+06} and RX J1037.5-5647 \citep{LaPalombara+09}. The HR is almost correlated (with a little delay) with the CR and varies of $\simeq \pm$ 10 \% around its average value along the pulse phase: this finding is rather common in the binary pulsars and reveals a limited energy dependence of the pulse profile. We also observed a slight increase in the pulsed fraction with the photon energy but, compared with other persistent BeXRBs \citep{LaPalombara+24}, in the case of \four\ the PF is much lower.

Since a simple absorbed PL model provided a rather poor fit of the source spectrum, we used various more elaborated models to describe it: on one hand, we considered two models with a single non-thermal component, either a partially covered PL or a CPL; on the other hand, we introduced a two-component model composed of a simple PL or CPL plus a thermal component, either a BB or an APEC model. All these models provide a good description of the source spectrum and none of them is clearly favoured compared with the other ones. For all models the estimated hydrogen column density is \nh $\sim (5-7) \times 10^{21}$ cm$^{-2}$, which is consistent with the values of interstellar absorption estimated with \RXTE\ and \Astro. For comparison, we note that the color excess of the optical counterpart is \textit{E(B-V)} = 0.75 \citep{Negueruela+96}, which implies an extinction $A_{\rm V}$ = 3.1 $\times$ \textit{E(B-V)} = 2.325; since the average relation between optical extinction and X--ray absorption is $N_{\rm H}$ = (2.87$\pm$0.12)$\times 10^{-22} A_{\rm V}$ cm$^{-2}$ \citep{Foight+16}, this implieas that $N_{\rm H}$ = (6.7$\pm$0.3)$\times10^{21}$ cm$^{-2}$, a value fully consistent with our result.

In the two spectral models with a PL component, the photon index $\Gamma$ is between 1.5 and 1.8, in agreement with the index values obtained in the cases of the \RXTE\ and \Astro\ spectra. Instead, with the two models with the CPL component we obtained best-fit values for $\Gamma$ and E$_{\rm cut}$ that are significantly lower than for the corresponding parameters obtained with \Astro. These differences are very likely due to the limited energy range covered by \XMM\ (0.2-12 keV). Therefore, it is possible that the power-law and cutoff components are not well constrained, particularly in the two-component models. Finally, regarding the partially absorbed PL model, we note that the inhomogeneous absorber medium described with the TBPCF component is very likely the clumpy polar wind coming from the Be companion star, which crosses the line of sight to the X-ray source \citep{Grinberg+20}.

Compared with a simple PL model, in the PL+BB model the additional thermal component improves significantly the spectral fit; it contributes for $\simeq$ 24 \% of the total X-ray flux and its temperature and radius are $kT_{\rm BB} \simeq$ 1.5 keV and $R_{\rm BB} \simeq$ 240 m, respectively. Therefore, for \four\ the properties of the detected BB excess are very similar to those observed in several other persistent BeXRBs: for most of these sources, in fact, this component has a temperature in the range 1-2 keV and a size of few hundred meters, and contributes for $\sim$ 20-40 \% to the total X-ray flux \citep{LaPalombara+24}. In the CPL+APEC model, instead, the additional thermal component provides only a marginal contribution to the total flux; moreover, the fit improvement obtained with the addition of the APEC component to the simple CPL is much lower than in the case of the BB addition to a simple PL.

In the \textit{EPIC} spectrum of \four\ we found no evidence of any emission or absorption feature, so we cannot confirm the presence of the broad Fe $K_{\alpha}$ line detected in the \Astro\ spectrum. On the contrary, our analysis provided an upper limit of $\lsim$ 100 eV on the equivalent width of a possible Fe line. It is very interesting to note that this result is fully aligned with what obtained for all the currently known persistent BeXRBs, since in none of them this type of feature has been observed up to now, and a comparable upper limit on its equivalent width has been estimated \citep{LaPalombara+24}.

We performed a phase-resolved spectral analysis in order to investigate the spectral dependence on the pulse phase. To this aim, we compared the spectra accumulated in two different phase intervals, corresponding to the maximum (H) and the minimum (S) of the HR. The independent fit of these two spectra showed that the source flux decreases of $\simeq$ 30 \% from phase H to phase S and that there is also a clear variability of the spectral parameters. For the PL+BB model, we found that the flux of both components is clearly variable and that the BB is significant at 99 \% c.l. in both spectra. Instead this is not true for the CPL+APEC model, since in both spectra the additional APEC component cannot be constrained. This result, together with the marginal APEC contribution to the time-average flux, disfavours the real presence of this type of component in the source spectrum. We also performed a simultaneous fit of the two phase-resolved spectra, in order to investigate in deeper detail the spectral variability. For the PL+BB model we verified that the spectral variability can be well described with a variation of only the BB flux, while this is not true in the case of a variation of only the PL flux; moreover, we verified that the two spectra are compatible with a constant BB component only if both the PL flux and photon index vary. Therefore, based on this results, the presence of a variable BB component is very likely.

The size of the BB emission region and the strong clues that it is variable suggest that it originates on the surface of the accreting NS, in agreement with the scenario proposed by \citet{Hickox+04} to explain the thermal excess in the low-luminosity HMXRBs. This hypothesis has been confirmed in most of the persistent BeXRBs, where it has been proven that the radius of the observed BB excess is consistent with the estimated radius of the accretion column onto the NS polar caps \citep{LaPalombara+24}. In order to verify if this hypothesis holds also in the case of \four, we calculated the column radius $R_{\rm col}$ and compared it with $R_{\rm BB}$. To this aim, for the accreting NS we assumed a mass $M_{\rm NS} = 1.4 M_{\odot}$, a radius $R_{\rm NS} = 10^6$ cm, and a magnetic field $B_{\rm NS} = 10^{12}$ G. Based on the estimated luminosity \lx(0.2-12 keV) $\simeq 1.3 \times 10^{35}$ \lum\ and on the relation $\dot M$ = \lx$R_{\rm NS}$/($GM_{\rm NS}$), we derived an accretion rate $\dot M \simeq 7 \times 10^{14}$ g s$^{-1}$ and a magnetospheric radius $R_{\rm m} \simeq 8.6 \times 10^8$ cm \citep{Campana+98}. Finally, we used the relation $R_{\rm col} \sim R_{\rm NS}$ ($R_{\rm NS}/R_{\rm m}$)$^{0.5}$ \citep{Hickox+04} to obtain $R_{\rm col} \simeq$ 340 m. This value is only $\sim$ 40 \% larger than the estimated BB radius $R_{\rm BB}$ = 240 m, which implies that the BB size is comparable with that of the NS polar caps. Therefore, it is very likely that they are the source of the observed BB component, and this hypothesis is further strengthened by the BB variability along the pulse phase.

	\section{Conclusions}
	\label{conclusions}

We have described the results obtained by analizing a $\simeq$ 33 ks \XMM\ observation of the BeXRB \four\ performed in April 2023. The source was detected at a luminosity level \lx $\simeq 9 \times 10^{34}$ \lum\ in the energy range 2-10 keV: this is the lowest level ever observed since the source discovery in the '70s, and it is also about 50 \% lower than the minimum luminosity previously observed. On the other hand, since this value is less than one order magnitude lower than the maximum luminosity observed in almost 50 years, this \XMM\ observation confirmed the limited flux variability of \four\ over long time scales.

We measured a pulse period $P_{\rm spin} = 103.301 \pm 0.005$ s, which implies a spin down $\pdot = 7.2 \times 10^{-10}$ s s$^{-1}$ as compared to the previous measurement in 2016. This moderate spin down is coherent with the low rate of matter transfer, via wind accretion, from the Be companion star onto the NS. The pulse shape shows a simple single-peak profile, with a moderate pulsed fraction at all energies.

The spectral analysis revealed the presence of a flux excess above the main power-law component, which can be described with a BB model with high temperature ($kT_{\rm BB} \simeq$ 1.5 keV) and small emission radius ($R_{\rm BB} \simeq$ 240 m); it contributes for $\sim$ 25 \% to the total flux and has a size comparable with that of the NS polar caps. Instead we found no evidence of emission features due to narrow ($\sigma <$ 0.5 keV) Fe $K_{\alpha}$ lines, with an upper limit of $\simeq$ 100 eV on the equivalent width of this type of feature.

We also performed a phase-resolved spectral analysis, which proved the spectral variability along the pulse phase. In this way we proved that the BB component is always present, and we also obtained strong clues that it is variable along the pulse phase. Therefore, we attributed its origin to the NS polar caps.

In summary, we verified that the timing and spectral properties of \four\ are fully consistent with those of most of the persistent BeXRBs. For this reason, we can consider it as a confirmed member of this class of sources.

\begin{acknowledgements}
NLP, LS, and PE acknowledge funding from INAF through the grants "Bando Ricerca Fondamentale INAF 2023" and "Bando Astrofisica Fondamentale 2024". PE acknowledges support from the PRIN MUR SEAWIND (2022Y2T94C), funded by the European Union - Next Generation EU, Mission 4 Component 1 CUP C53D23001330006,
and the INAF Grant BLOSSOM.
\end{acknowledgements}

\bibliographystyle{aa}
\bibliography{biblio}

\end{document}